\documentclass[11pt]{article}
\usepackage{latexsym}

\def\be{\begin{equation}}
\def\ee{\end{equation}}
\def\bea{\begin{eqnarray}}
\def\eea{\end{eqnarray}}
\def\pa{\partial}

\def\fn{\footnote}

\def\case#1/#2{\textstyle\frac{#1}{#2}}

\begin{document}

\vspace{.5in}

\begin{center}
\Large\bf
{GEOMETRY, EFE'S AND FOUNDATIONS FOR BRANEWORLDS}\normalfont
\\
\vspace{.5in} \normalsize \large{Edward Anderson}
\\
\normalsize
\vspace{.4in}
{\em  Astronomy Unit, School of Mathematical Sciences, \\
Queen Mary, University of London E1 4NS, U.K. }

\vspace{.2in}
\end{center}
\vspace{.3in}

\noindent I draw on GR Cauchy problem (CP) and initial value problem (IVP) mathematics \cite{CBY} 
to make a number of points about Shiromizu--Maeda--Sasaki (SMS) \cite{SMS} 5-d Einstein field 
equation-type (EFE) braneworlds.  

One issue \cite{ATlett} is why SMS chose their particular steps in  their formulation of 
braneworld equations explicitly in terms of Weyl tensor projections.  For in the GR CP, 
similar steps are most commonly used to eliminate the Weyl tensor projections.  Although 
other GR CP formulations are explicitly in terms of Weyl tensor projections, the specific 
reasons for these formulations do not extend to the braneworld application.  There, rather, 
explicit formulation in terms of Weyl tensor projections is convenient for those who wish to 
(partly) set these projections to zero.  This is not physically justifiable if these 
braneworlds are to be interpreted within 5-d GR.  Another issue is whether this actually bears 
any rigorous relation whatsoever with extracting predictions from string theory.  

Given a 5-d EFE context however, surely one should judge it by GR's dynamical content.  
The presence of a brane (a thin 3+1 hypersurface privileged by having tension and matter pinned 
to it) makes this a new and difficult problem.  From the GR CP it is suggestive that the 
components of the EFE's on a hypersurface are a means of determining the nature of nearby 
hypersurfaces (and thus building up the higher-dimensional spacetime), rather than being some 
analogue of the EFE's (which is in any case still implicitly such a means, through the 
equations required to close the system).    

But I do not favour the idea of interpreting the 10 on-brane field equations as a means of 
determining the surrounding bulk.  For \cite{ATpap}, this does not make good causal sense and is 
not known to depend continuously on the on-brane data.  I rather favour extending the space 
corresponding to a brane snapshot to make a bulk snapshot.  This is a standard IVP, to be followed by 
a standard CP to see what happens to the brane and bulk at later times.  This method {\it is} 
causally sensible and follows in the tradition of continuously-dependant problems.  It would 
ultimately permit study of whether branes are stable given GR dynamics.  I envisage the 
evolution part of this problem to be hard however, and restrict attention below to the data 
problem.  This by itself has the smaller merit of determining possible surrounding bulk shapes 
for on-brane compact objects (the cigars versus pancakes question).  

The equations to solve are 
\be
K^2 - \frac{3}{4}K_{ij}^{\mbox{\scriptsize T\normalsize}}K^{ij\mbox{\scriptsize T\normalsize}} + R = 2\rho 
\mbox{ } , \mbox{ }
\label{Gauss}
\ee
\be
D_jK_i^{\mbox{\scriptsize T\normalsize}j} - \frac{3}{4}D_iK = j_i
\ee
(where $K_{ij}^{\mbox{\scriptsize T\normalsize}}$ is the 4-space tracefree extrinsic curvature, 
$R$ the 4-space Ricci scalar, $D_i$ the 4-space covariant derivative and $\rho$, $j_i$ are 
the obvious 4-space projections of 5-spacetime energy-momentum).  These conveniently decouple 
if approached by York's method \cite{CBY}.  For $j_i = 0$ it suffices to have  
$K_{ij}^{\mbox{\scriptsize T\normalsize}}$ transverse, and then solve the 4-d 
free-$\alpha$ {\it Lichnerowicz--York equation } [a conformalized version of (\ref{Gauss})], 
\be
D^2\phi = \frac{\phi}{6}
\left(
R - K_{ij}^{\mbox{\scriptsize T\normalsize}}K^{ij\mbox{\scriptsize T\normalsize}}\phi^{-6} + K^2\phi^2 - 2\rho\phi^{\alpha}
\right) 
\ee
for the conformal factor $\phi$.  The good behaviour of this equation depends on the signs 
of the polynomial's coefficients.  Of note, a desire for AdS-like bulks leads to $\rho < 0$,  
opposite to theoretical numerical relativity.  

The main differences between the above IVP and standard ones however lies in the associated 
boundary conditions (b.c's) \cite{ATpap}.  Let's work with a ``$S^2 \times \Re$" metric 
$ds^2 = dz^2 + e^{w(z, r)}dx_{\gamma}dx^{\gamma}$ to consider a compact 
object on the brane (where the $\gamma$ run over the brane and $z \geq 1$ extends into the bulk).  
From the $Z_2$ junction condition, the on-brane b.c is 
\be
\left. \left[\frac{\pa \phi}{\pa z} + \frac{\phi}{2}\left(   \frac{\pa w}{\pa z} - \frac{\kappa_5^2\tilde{\lambda}}{2}\phi \right)\right] \right| _{z = 1} = 0 
\label{nonlinbc}
\ee
which is unusual in being nonlinear\fn{If $w$ is taken to be the unknown \cite{SSNN}, the b.c is  
linear, but unlike here, that method is not known to be simply exendible away from the simplest cases.} 
(but nevertheless of a tractable type).  The notion of far from the brane is itself not 
standard GR asymptotics but rather a troublesome notion (by typical caustic formation).  In any 
case one expects a Dirichlet b.c.  The outer-radius b.c is expected to be asymptotically flat 
on-brane but is less clear off-brane.  For a star one may use an inner-radius Neumann b.c.  
But this is not acceptable for a black hole since it would entail  
on-singularity data prescription.  One would usually use inversion-in-$S^2$ within the apparent 
horizon giving a Robin b.c.  But now we have the added problem of not knowing in advance neither 
the horizon shape nor how far the singularity protrudes into the bulk.  One may need to `shoot' 
if one wishes to avoid both on-singularity prescription and excision of causally-connected regions.  

\mbox{ }

\noindent{\bf Acknowledgements }

\mbox{ }

I would like to thank Reza Tavakol for working with me on these issues.

\end{document}